\documentclass[pra,twocolumn,groupedaddress,showpacs,floatfix]{revtex4}
\usepackage{graphics}
\usepackage{graphicx}
\usepackage{bm}
\usepackage{amsmath}
\usepackage{amsfonts}
\usepackage{amssymb}
\usepackage{latexsym}
\usepackage{color}
\usepackage{bbold}
\usepackage{braket}
\def\Tr{\mbox{Tr}}

\begin{document}

\title{Quantum-enhanced metrology with the single-mode coherent states \\ of an optical cavity inside a quantum feedback loop}
\author{Lewis A. Clark, Adam Stokes, and Almut Beige}
\affiliation{The School of Physics and Astronomy, University of Leeds, Leeds LS2 9JT, United Kingdom}
\date{\today}

\begin{abstract}
In this paper, we use the non-linear dynamics of the individual quantum trajectories of an optical cavity inside an instantaneous quantum feedback loop to measure the phase shift between two pathways of light with an accuracy above the standard quantum limit. The feedback laser provides a reference frame and constantly increases the dependence of the state of the resonator on the unknown phase. Since our quantum metrology scheme can be implemented with current technology and does not require highly-efficient single photon detectors, it should be of practical interest until highly-entangled many-photon states become more readily available.
\end{abstract}

\maketitle

\section{Introduction}

In general, there are two main strategies for reducing the uncertainty in an experimentally measured quantity. One method is to repeat the experiment many times.  Another is to use more of an appropriate resource, $N$, in every run of the experiment. However, increasing $N$ is not always possible. Suppose we want to measure the phase shift $\varphi$ caused by a delicate material with the help of a standard light interference experiment. Increasing the number of photons passing through can increase the accuracy of every phase measurement but also limits the lifetime of the sample \cite{lifetime1,lifetime2,lifetime3}. In this case, it is important that every run of the experiment is as accurate as possible. To allow for a fair comparison of different measurement schemes, the error propagation formula 
\begin{eqnarray} \label{Error}
\Delta \varphi &=& \frac{\Delta M}{\left| {{\partial M} \over {\partial \varphi}} \right|} 
\end{eqnarray}
can be used to calculate the accuracy $\Delta \varphi$ of a given signal $M(\varphi)$ \cite{review}. Here $\Delta M$ denotes the uncertainty (or resolution) of $M$, while the visibility, $| {\partial M} / {\partial \varphi}|$, tells us how sensitive $M$ is to changes in $\varphi$. 

\begin{figure}[t]
\centering
\includegraphics[width=0.40\textwidth]{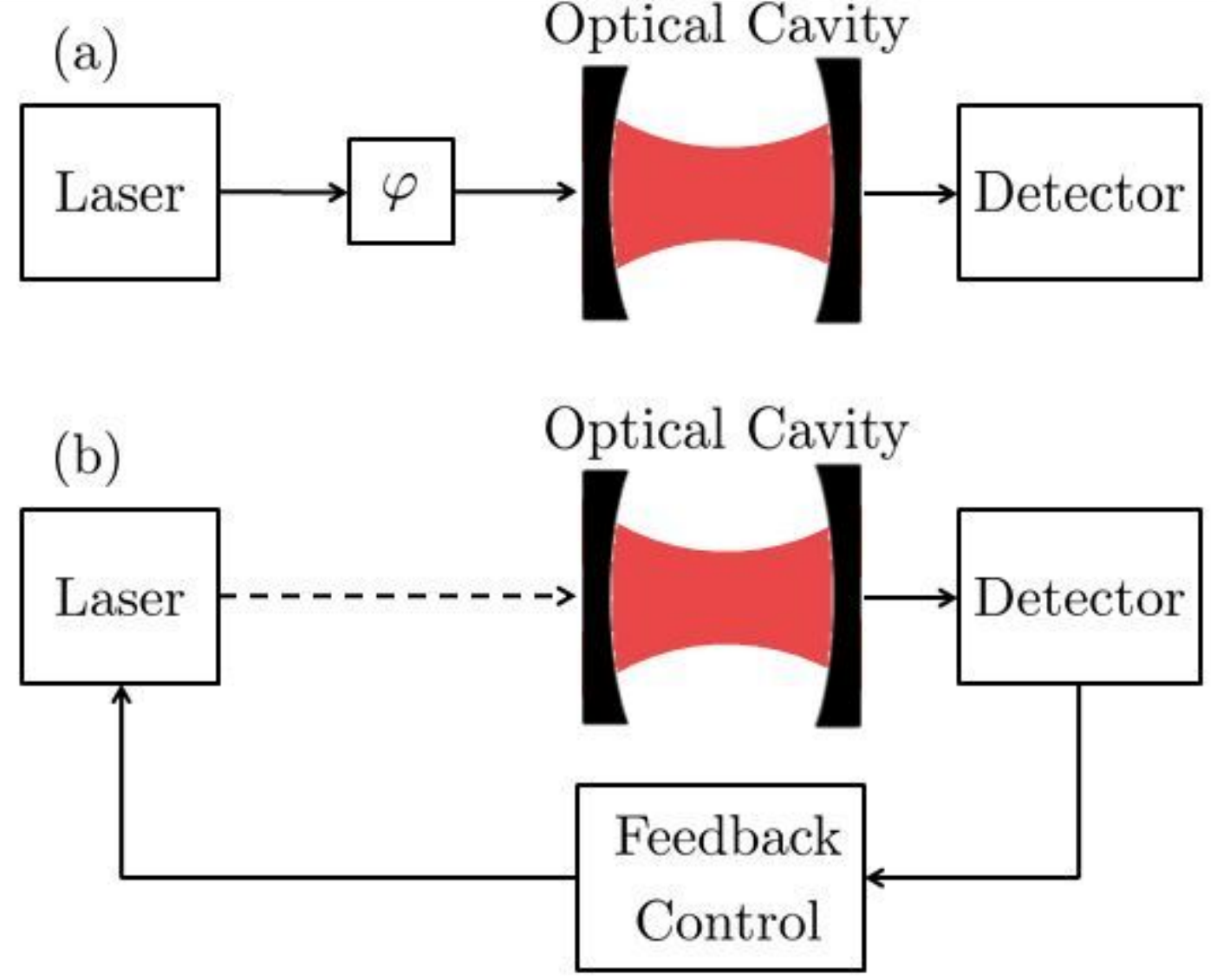}
\caption{[Color online] The proposed quantum-enhanced metrology scheme involves two main stages. (a) During the preparation stage, a laser experiences an unknown phase $\varphi$ before entering the resonator, thereby preparing the cavity in a coherent state $|\alpha \rangle$ with $\alpha$ as in Eq.~(\ref{polar}). (b) During the measurement stage, the continuous laser driving is replaced by an instantaneous feedback loop. Whenever a photon is detected, with a finite detector efficiency $\eta$, the feedback laser displaces the resonator field. Whether or not the feedback pulse increases the energy inside the cavity and how often it is triggered depends strongly on $\varphi$.} \label{Cavity-Prep}
\end{figure}

Using $N$ independent photons, the scaling of the lower bound of the uncertainty of the phase measurement between two pathways of light, $\Delta \varphi_{\rm class}$, is given by the standard quantum limit,
\begin{eqnarray} \label{Limitsclass}
\Delta \varphi_{\rm class} &\propto & N^{-0.5} \, .
\end{eqnarray}
There are different ways in which this scaling can be improved. One way is to expose the incoming photons to a 
non-linear Hamiltonian \cite{non-lin}. In this case, the uncertainty of a single phase measurement, $\Delta \varphi_{\rm non-lin}$, scales as
\begin{eqnarray} \label{Limitsnonlin}
\Delta \varphi_{\rm non-lin} & \propto & N^{- 0.5 \, k} \, ,
\end{eqnarray}
where $k$ denotes the order of the present non-linearity. However, highly-efficient optical non-linearities are hard to implement in general. Another way to obtain an enhancement is to replace the incoming independent photons by entangled ones \cite{Q.Met, Q.Met2,Kok,Berry}. Using entanglement, the measurement uncertainty, $\Delta \varphi_{\rm quant}$, can be as low as the Heisenberg limit,  
\begin{eqnarray} \label{Limitsquant}
\Delta \varphi_{\rm quant} &\propto & N^{-1} \, .
\end{eqnarray}
To extract information from highly non-classical  photon states \cite{Caves,Squeeze,N00Ncreation,Gerry0,Gerry4,Paul,ECS,ECS2,M&M,mm'}, quantum metrology schemes may use techniques such as quantum feedback, photon parity measurements, probes with fluctuating number states and photon subtraction \cite{Berry2,Gerry-1,Gerry,Fluctuation,Gerry3}. Although it is possible to realise multi-photon entanglement in the laboratory \cite{LIGO}, quantum metrology has not yet become readily-available for a wide range of applications. 

To overcome this problem, this paper proposes to measure the unknown phase shift $\varphi$ between two pathways of light using a leaky optical resonator inside an instantaneous quantum feedback loop. As illustrated in Fig.~\ref{Cavity-Prep}, the quantum-enhanced metrology scheme that we propose here consists of two main stages. Firstly, the preparation stage prepares the cavity field  in a coherent state $|\alpha \rangle$ with
\begin{eqnarray} \label{polar}
\alpha &=& |\alpha| \, {\rm e}^{{\rm i} \varphi} \, .
\end{eqnarray} 
Afterwards, during the measurement stage, the cavity is placed inside a quantum feedback loop. Whenever a photon is detected, a laser pulse is applied, which does not experience the unknown phase $\varphi$. The pulse displaces the field inside the resonator in a certain direction, thereby providing the reference frame for the proposed phase measurement. For the feedback pulse to be approximately instantaneous, it needs to be short compared to the average cavity photon life time $1/\kappa$. In the following we extract information about the unknown phase $\varphi$ from the temporal quantum correlations in the spontaneous photon emissions of the optical resonator. The measurement of these correlations does not require highly-efficient single photon detectors. Hence realising the experimental setup in Fig.~\ref{Cavity-Prep} is feasible with current technology \cite{Kuhn,Kimble,Laser}. 

As we shall see below, the only density matrix $\rho$ of the cavity field with a vanishing time derivative $\dot \rho = 0$ is the vacuum state. When starting in this state, the system remains there and never experiences a feedback pulse. However, in general, the cavity field remains in a single-mode coherent state $|\alpha \rangle$ with $\alpha \neq 0$. In many cases, $\alpha$ increases rapidly in time. Unlike most quantum optical systems with spontaneous photon emission, the resonator {\em never} reaches a stationary state \cite{Masalov,students}. The final state of the cavity depends very strongly on the phase $\varphi$, which has initially been imprinted onto the resonator (c.f.~Eq.~(\ref{polar})). Moreover, the temporal quantum correlations of the single trajectories of the cavity field cannot be expressed as first-order expectation values and do not evolve according to a set of linear differential equations. Their {\em non-linear} dynamics is what allows us to perform better-than-classical phase estimation. Using the dissipative dynamics of open quantum systems \cite{Breuer-Petruccione,Wiseman-Milburn,Adam,Clark}, Refs.~\cite{Braun,Openmet} already designed quantum metrology schemes that exceed the standard quantum limit. The main advantage of the scheme that we discuss here is that it is easy to realise experimentally. Our quantum-enhanced metrology scheme should be of practical interest until highly-entangled many-photon states become more readily available.

Temporal quantum correlations \cite{Kok2,Kok3} and sequential measurements \cite{Ref1,Ref2,Ref3} in open quantum systems are known to constitute an interesting resource for technological applications. To illustrate this, we show in the following that subsequent measurements on a single quantum system are in general equivalent to single-shot measurements on an entangled state of several systems. Suppose a two-dimensional quantum system is in an initial state $|\psi \rangle$ and subsequent generalised measurements are performed, which can be described by two Kraus operators $K_0$ and $K_1$ of the form
\begin{eqnarray} \label{Kraus2}
K_i &=& |\tilde \xi_i \rangle \langle \xi_i | \, .
\end{eqnarray}
Here $|\xi_0 \rangle$ and $|\xi_1 \rangle$ are two orthogonal states with $\langle \xi_0 |\xi_1 \rangle = 0$. However no such constraint is imposed on the tilde-states $|\tilde \xi_0 \rangle$ and $|\tilde \xi_1 \rangle$ \cite{Kraus}. In case of two measurements, the initial state of the system changes according to 
\begin{eqnarray}\label{psi}
|\psi \rangle &\rightarrow & \left\{ \begin{array}{l} K_0 \, |\psi \rangle \rightarrow \left\{ \begin{array}{l} K_0 K_0 \, |\psi \rangle \\ K_1 K_0 \, |\psi \rangle \end{array} \right. \\[0.4cm] K_1 \, |\psi \rangle \rightarrow \left\{ \begin{array}{l} K_0 K_1 \, |\psi \rangle \\ K_1 K_1 \, |\psi \rangle \end{array} \right. \end{array} \right. 
\end{eqnarray}
up to normalisation factors, which we neglect here for simplicity. Moreover suppose we perform a single-shot measurement of $K_0$ and $K_1$ on two quantum systems prepared in an effective state $|\psi_{\rm eff} \rangle$,
\begin{eqnarray} \label{ent}
|\psi_{\rm eff} \rangle &=& \sqrt{p_{00}} \, |\xi_0 \rangle \otimes |\xi_0 \rangle + \sqrt{p_{01}} \, |\xi_0 \rangle \otimes |\xi_1 \rangle \nonumber \\
&& + \sqrt{p_{10}} \, |\xi_1 \rangle \otimes  |\xi_0 \rangle  + \sqrt{p_{11}} \, |\xi_1 \rangle \otimes  |\xi_1 \rangle 
\end{eqnarray}
with the coefficients $p_{ij}$ equal to
\begin{eqnarray}
p_{ij} &=& \| K_j K_i \, |\psi \rangle \|^2 \, .
\end{eqnarray} 
It is easily to see that both measurements yield the outcome ``$ij$'' with exactly the same probability. This means, the states $|\psi \rangle$ and $|\psi_{\rm eff} \rangle$ have the same information content. However, $|\psi_{\rm eff} \rangle$ is in general an entangled state. For example, if $K_0 = |\xi_1 \rangle \langle \xi_0 |$ and $K_1 = |\xi_0 \rangle \langle \xi_1 |$, then $|\psi_{\rm eff} \rangle = \sqrt{p_{01}} \, |\xi_0 \rangle \otimes |\xi_1 \rangle + \sqrt{p_{10}} \, |\xi_1 \rangle \otimes  |\xi_0 \rangle$, which can be maximally entangled. Analogously, one can show that $N$ successive measurements on a single system are in general equivalent to a single-shot measurement of $N$ entangled quantum systems. This fact can be exploited for quantum metrology when using Kraus operators that depend on the unknown parameter.

The quantum-enhanced metrology scheme that we propose here extracts information about the unknown phase $\varphi$ of the initial state in Eq.~(\ref{polar}) by performing $N$ successive measurements on a single quantum system. This means, our scheme is equivalent to performing single-shot measurements on a combination of $N$ entangled quantum systems. Instead of multi-partite entanglement, we use temporal quantum correlations \cite{temp,temp2}. The main resource of our quantum metrology scheme, i.e.~the number of queries posed during each run of the experiment, hence equals the number of successive measurements on the cavity field. In other words, it essentially equals the number of time steps in which the cavity either emits a photon or not, which is proportional to the duration of the proposed experiment. As long as the behaviour of the cavity field generates temporal quantum correlations with non-linear dynamics, actual physical entanglement does not need to be present \cite{non-lin,inwrap}.  We are therefore not in contradiction with previous work that claims entanglement is required to go beyond standard scaling, as in such cases only linear generators of change in the unknown parameter are considered \cite{Q.Met,Q.Met2}. 

There are five sections in this paper. In Section \ref{Master Equations}, we discuss how to model an open quantum system inside an instantaneous feedback loop, thereby providing the general theoretical background for our work. In Section \ref{Cavity} we analyse the dynamics of a laser-driven optical cavity with instantaneous quantum feedback in the form of very short strong laser pulses. In Section \ref{Metrology}, we design a quantum-enhanced metrology scheme with single-mode coherent states. We then calculate its accuracy with respect to intensity measurements and with respect to second-order photon correlation measurements. Finally we summarise our findings in Section \ref{Conclusion}.

\section{Quantum optical master equations with instantaneous feedback} \label{Master Equations}

In this section, we give a brief introduction to the modelling of open quantum systems \cite{Breuer-Petruccione,Adam}. To do so, we consider a general quantum system which interacts with a surrounding bath. This bath is assumed to also interact with an external environment, which causes it to thermalise. This means, the environment constantly resets the bath into its environmentally preferred state -- its so-called pointer state \cite{Einselection}. The resulting effective time evolution of the open quantum system is approximately Markovian and its density matrix $\rho_{\rm S}$ obeys a master equation in Lindblad form. Since the bath surrounding the quantum system is continuously monitored by the environment for the detection of spontaneously emitted photons \cite{Reset,Molmer,Carmichael}, this master equation can be unravelled into an infinite set of physically-meaningful quantum trajectories. Considering such an unravelling and assuming that the instantaneous feedback is triggered by sudden changes of the state of the quantum system, it becomes clear how to incorporate instantaneous feedback into the master equation \cite{Wiseman-Milburn,Clark}. 

\subsection{Master equations without feedback} 

Let us first have a closer look at an open quantum system without feedback. 

\subsubsection{Hamiltonian of system and bath}

The Hamiltonian $H$ of such a system and its surrounding bath can be split it into two parts,
\begin{eqnarray} \label{H}
H &=& H_0+ H_1 
\end{eqnarray}
with $H_0$ denoting the free energy of the quantum system and its bath,
\begin{eqnarray} \label{H2}
H_0 &=& H_{\rm S}+H_{\rm B} \, ,
\end{eqnarray}
and with $H_1$ consisting of two terms,
\begin{eqnarray} \label{H22}
H_1 &=& H_{\rm int} + H_{\rm SB} \, . 
\end{eqnarray}
Here $H_{\rm SB}$ describes system-bath interactions and $H_{\rm int}$ describes the internal system dynamics. Moving into the interaction picture with respect to $H_0$, the Hamiltonian simplifies to interaction Hamiltonian $H_{\rm I}(t) = U_0^\dagger (t ,0) \, H_1 \,  U_0(t ,0)$, which is of the general form
\begin{eqnarray} \label{H3}
H_{\rm I}(t) &=& H_{\rm int \, I} + H_{\rm SB \, I} \, .
\end{eqnarray}
In the following, we use this Hamiltonian to derive a Markovian master equation for open quantum systems. 

\subsubsection{Environmental effects}

Suppose the state of the quantum system at time $t$ is given by the density matrix $\rho_{\rm S} (t)$. Moreover, adopting the ideas of Refs.~\cite{Einselection,Reset,Molmer,Carmichael,Adam}, we assume in the following that the bath surrounding the quantum system is in general in its environmentally preferred state -- the so-called einselected state or pointer state -- which we denote by $|0 \rangle$. Hence the general density matrix of system and bath at some time $t$ can be written as
\begin{eqnarray} \label{Evolution23}
\rho_{\rm SB} (t) &=& \left\vert 0 \right\rangle \rho_{\rm S}(t) \left\langle 0 \right\vert \, .
\end{eqnarray}
As argued in Ref.~\cite{Einselection}, the pointer state $|0\rangle$ is environmentally preferred because it minimises the entropy of the bath. Hence the bath only evolves due to system-bath interactions but is invariant with respect to its own internal dynamics. 

Next, we assume that system-bath interactions perturb the state of the bath on a time scale $\Delta t$, which is short compared to the time scale given by the effective internal dynamics of the quantum system. During this time interval, the density matrix $\rho_{\rm SB} (t)$ evolves via the time evolution operator $U_{\rm I} (t+\Delta t , t)$ into a new density matrix $\rho_{\rm SB} (t+\Delta t)$ given by
\begin{eqnarray} \label{Evolution2}
\rho_{\rm SB} (t + \Delta t) &=& U_{\rm I}(t+\Delta t , t) \left\vert 0 \right\rangle \rho_{\rm S}(t) \left\langle 0 \right\vert U_{\rm I}^{\dagger} \left(t+\Delta t , t\right) . \nonumber \\
\end{eqnarray}
Following the discussion in Refs.~\cite{Einselection,Reset,Adam}, we now assume that environmental interactions subsequently relax the reservoir very rapidly back into its environmentally preferred state. If the environment acts only locally and does not affect the expectation values of the quantum system, the result of this thermalisation is a new system-bath density matrix
\begin{eqnarray} \label{Evolution3}
\rho_{\rm SB} (t + \Delta t) &=& \left\vert 0 \right\rangle \rho_{\rm S} (t + \Delta t) \left\langle 0 \right\vert 
\end{eqnarray}
with the state of the system given by
\begin{eqnarray}
\rho_{\rm S} (t + \Delta t) &=& \Tr_{\rm B} \left( \rho_{\rm SB} (t + \Delta t) \right) \, .
\label{General evolution no feedback}
\end{eqnarray}
Effectively, only $\rho_{\rm S}(t)$ has evolved over the interval $\Delta t$, and its dynamics can be summarised by the master equation 
\begin{eqnarray}
\dot{\rho}_{\rm S} (t) &=& \frac{1}{\Delta t} \, \left[ \rho_{\rm S} \left(t + \Delta t \right) - \rho_{\rm S}\left( t \right) \right] \, .
\label{Derivative}
\end{eqnarray}

\subsubsection{Perturbative Expansions}

Given a clear time scale separation between the effective inner dynamics of the quantum system and the relevant system-bath interactions, the right-hand-side of Eq.~(\ref{Derivative}) can be evaluated using second-order perturbation theory. To do so, we write the time evolution operator $U_{\rm I} (t+\Delta t ,t) $ as
\begin{eqnarray}
U_{\rm I} (t+\Delta t ,t) &=& 1 - \frac{i}{\hbar} \int_t^{t+\Delta t} {\rm d}t' \, H_{\rm I} (t') \nonumber \\
&& \hspace*{-0.9cm} - \frac{1}{\hbar^2} \int_t^{t+\Delta t} {\rm d}t' \int_t^{t'} {\rm d}t'' \, H_{\rm I}(t') H_{\rm I} (t'') \, . ~~~~
\end{eqnarray}
Substituting this equation into Eq.~(\ref{Evolution2}) and combining the result with Eqs.~(\ref{General evolution no feedback}) and (\ref{Derivative}), we find that 
\begin{widetext}
\begin{eqnarray} \label{master}
\dot{\rho}_{\rm S} (t) &=& - \frac{\rm i}{\hbar} \big[ \, H_{\rm int \, I} (t) , \rho_{\rm S} (t) \, \big] - {1 \over \Delta t} \, \frac{1}{\hbar^2} \,\int_t^{t+\Delta t} {\rm d}t' \int_t^{t'} {\rm d}t'' \, \Big( \langle 0| H_{\rm SB \, I} (t') H_{\rm SB \, I} (t'') |0 \rangle \, \rho_{\rm S} (t) + {\rm H.c.} \Big) \nonumber \\
&& + {1 \over \Delta t} \, \frac{1}{\hbar^2} \, \Tr_{\rm B} \left( \int_t^{t+\Delta t} {\rm d}t' \int_t^{t+\Delta t} {\rm d}t'' \, H_{\rm SB \, I}(t') \left\vert 0 \right\rangle \rho_{\rm S} (t) \left\langle 0 \right\vert H_{\rm SB \, I}\left(t'' \right) \right) 
\end{eqnarray}
\end{widetext}
up to zeroth order in $\Delta t$. When deriving this equation, it has been taken into account that $\Delta t$ is relatively small and that a typical bath has infinitely many degrees of freedom. Therefore, the double integrals in Eq.~(\ref{master}) scale in general as $\Delta t$, and not as $\Delta t^2$. 

\subsection{Unravelling into quantum trajectories} \label{unravel}

To incorporate instantaneous feedback \cite{Wiseman-Milburn,Clark} into the above master equation, we notice that the application of feedback requires monitoring the bath for triggering signals. Assuming the presence of such measurements on the above introduced time scale $\Delta t$ allows us to unravel the above master equation into physically meaningful quantum trajectories \cite{Reset,Molmer,Carmichael}. Denoting the (unnormalised) density matrix of the subensemble of quantum systems for which the bath remains in its environmentally preferred state $|0 \rangle$ by $\rho_{\rm S}^0 (t)$, and the (unnormalised) density matrix of the subensemble for which the bath changes by $\rho_{\rm S}^{\neq} (t)$, one can show that 
\begin{eqnarray} \label{Evolution44}
\dot{\rho}_{\rm S} (t) &=& \dot{\rho}_{\rm S}^0 (t) + \dot{\rho}_{\rm S}^{\neq} (t) 
\end{eqnarray}
with
\begin{eqnarray} \label{master2}
\dot{\rho}_{\rm S}^0 (t) &=& - \frac{\rm i}{\hbar} \, \big[ \, H_{\rm int \, I} (t') , \rho_{\rm S} (t) \, \big] \nonumber \\
&& - {1 \over \Delta t} \, \frac{1}{\hbar^2} \int_t^{t+\Delta t} {\rm d}t' \int_t^{t'} {\rm d}t''  \nonumber \\
&& \times \Big( \langle 0| H_{\rm SB \, I} (t') H_{\rm SB \, I} (t'') |0 \rangle \, \rho_{\rm S} (t) + {\rm H.c.} \Big)  ~~~
\end{eqnarray}
and 
\begin{eqnarray}\label{master38}
\dot{\rho}^{\neq}_{\rm S}(t) & = & {1 \over \Delta t} \, \frac{1}{\hbar^2} \, \Tr_{\rm B} \Bigg( \int_t^{t+\Delta t} {\rm d}t' \int_t^{t+\Delta t} {\rm d}t'' \nonumber \\
&& \times H_{\rm SB \, I}(t') \left\vert 0 \right\rangle \rho_{\rm S} (t) \left\langle 0 \right\vert H_{\rm SB \, I}\left(t'' \right) \Bigg) \, .
\end{eqnarray}
Notice that the trace operation in Eq.~(\ref{General evolution no feedback}) is independent of the basis in which it is performed. Consequently, the dynamics of $\rho_{\rm S}$ does not depend on how the bath is actually measured. 

For very small $\Delta t$, Eq. (\ref{master2}) can be written in the more compact form
\begin{eqnarray} \label{master22}
\dot{\rho}_{\rm S}^0 (t) &=& - \frac{\rm i}{\hbar} \big[ H_{\rm cond} (t) \, \rho_{\rm S} (t) - \rho_{\rm S} (t) \, H_{\rm cond}^\dagger (t) \big]  
\end{eqnarray}
with $H_{\rm cond} (t)$ being the (non-Hermitian) conditional Hamiltonian of the open quantum system. This means, $\rho_{\rm S}^0 (t) $ evolves effectively according to a Schr\"odinger equation. If the quantum system is initially in a pure state $|\psi_{\rm S} (t) \rangle$, it remains pure as long as the state of the bath does not change due to system-bath interactions \cite{Reset,Adam}. The probability for the bath to remain in its preferred state $|0\rangle$ for a time $\Delta t$ equals
\begin{eqnarray} \label{master3333}
P_0(\Delta t) &=& \| \, U_{\rm cond} (t+\Delta t,t) \, |\psi_{\rm S} (t) \rangle \, \|^2 \nonumber \\
&=& {\rm Tr} \left( \rho_{\rm S}^0 (t+\Delta t) \right) \, , 
\end{eqnarray}
where $U_{\rm cond} (t+\Delta t,t) $ denotes the time evolution operator corresponding to $H_{\rm cond}(t)$. 

\subsection{Master equations with instantaneous feedback}

Repeating the above derivation of Eq.~(\ref{master}) while assuming that the quantum system experiences a unitary feedback operation, $R_m$, with probability $\eta_m$ whenever the state of the bath is found in $|m \rangle$ and $m \neq 0$, we arrive again at Eqs.~(\ref{master2}) and (\ref{master22}) but with Eq.~(\ref{master38}) replaced by
\begin{widetext}
\begin{eqnarray} \label{masterx}
\dot{\rho_{\rm S}}^{\neq} (t) &=& \frac{1}{\Delta t} \, \frac{1}{\hbar^2} \, \sum_{m \neq 0} (1 - \eta_m) \int_t^{t+\Delta t} {\rm d}t' \int_t^{t+\Delta t} {\rm d}t'' \, \langle m | \, H_{\rm SB \, I}(t') \left\vert 0 \right\rangle \rho_{\rm S} (t) \left\langle 0 \right\vert H_{\rm SB \, I}\left(t'' \right) |m \rangle \nonumber \\
&& + \frac{1}{\Delta t} \, \frac{1}{\hbar^2} \, \sum_{m \neq 0} \eta_m \, \int_t^{t+\Delta t} {\rm d}t' \int_t^{t+\Delta t} {\rm d}t'' \, R_m \, \langle m | \, H_{\rm SB \, I}(t') \left\vert 0 \right\rangle \rho_{\rm S} (t) \left\langle 0 \right\vert H_{\rm SB \, I}\left(t'' \right) |m \rangle \, R_m^\dagger \, .
\end{eqnarray}
\end{widetext}
The kind of feedback described in this subsection is often referred to as {\em instantaneous feedback}, since it acts on the time scale $\Delta t$ which is much shorter than the time scale given by the internal system dynamics \cite{Wiseman-Milburn}. 

\section{An optical cavity inside a quantum feedback loop} \label{Cavity}

The experimental setup that we consider in this paper is shown in Fig.~\ref{Cavity-Prep}. It contains a laser-driven optical cavity, a photon detector with a finite efficiency $\eta$ and a quantum feedback loop. In this section, we build on the results of the previous section to obtain the relevant equations for the dynamics of the cavity field, with and without feedback.

\subsection{The relevant Hamiltonians}

The Hamiltonian $H_0$ in Eq.~(\ref{H2}) contains two contributions, $H_{\rm S}$ and $H_{\rm B}$. For the experimental setup in Fig.~\ref{Cavity-Prep}, $H_{\rm S}$ describes the free energy of the optical cavity,
\begin{eqnarray}
H_{\rm S} & = & \hbar \omega_{\rm cav} \, c^{\dagger} c \, ,  
\end{eqnarray}
where $\hbar \omega_{\rm cav}$ denotes the energy of a single photon and $c$ and $c^\dagger$ are bosonic photon annihilation and creation operators with $[c,c^\dagger] = 1$. The Hamiltonian $H_{\rm B}$ represents the free energy of the surrounding bath modes, the free radiation field.  As usual in quantum optics, we have
\begin{eqnarray} \label{HB2}
H_{\rm B} &=& \sum_{{\bf k} \lambda} \hbar \omega_{{\bf k} \lambda}  \, a_{{\bf k} \lambda} ^\dagger a_{{\bf k} \lambda}  \, ,
\end{eqnarray}
where $a_{{\bf k} \lambda} $ denotes the annihilation operator of a single photon with frequency $\omega_k$, wave vector ${\bf k}$ and polarisation $\lambda$, while $a_{{\bf k} \lambda}^\dagger $ denotes the corresponding creation operator with $[a_{{\bf k} \lambda}, a_{{\bf k}' \lambda'}^\dagger] = \delta_{\lambda\lambda'}\delta_{{\bf k k}'}$. In addition, we need to specify the Hamiltonian for the internal dynamics of the system and the system-bath interaction, $H_{\rm int}$ and $H_{\rm SB} $ in Eq.~(\ref{H22}). Going straight into the interaction picture with respect to $H_0$ and applying the usual rotating wave approximation, we find that 
\begin{eqnarray} \label{Cavity interaction Hamiltonian}
H_{\rm int \, I} &=& {\textstyle {1 \over 2}} \hbar  \Omega \, \left( {\rm e}^{{\rm i} \varphi} \, c +  {\rm e}^{- {\rm i} \varphi}  \, c^\dagger \right) \, , \nonumber \\
H_{\rm SB \, I} &=&\sum_{{\bf k} \lambda} \hbar g_{{\bf k} \lambda}  \, a_{{\bf k} \lambda} ^\dagger c + {\rm H.c.} 
\end{eqnarray}
The first Hamiltonian describes the resonant driving of the cavity by an external laser field with Rabi frequency $\Omega \, {\rm e}^{{\rm i} \varphi}$. Here $\Omega$ is assumed to be real, while $\varphi$ specifies the phase of the laser. Moreover, $H_{\rm SB \, I}$ models the exchange of photon excitation between the cavity and the free radiation field with $g_{{\bf k} \lambda} $ denoting the respective coupling constants.

\subsection{The relevant master equations}

Now that we have identified all the relevant Hamiltonians for the experimental setup in Fig.~\ref{Cavity-Prep}, we can substitute them into Eq.~(\ref{master}). Calculating the respective integrals and absorbing level shifts into the free energy term $H_0$, we find that the master equation of a laser-driven optical cavity without feedback equals
\begin{eqnarray} \label{Cavity master equation}
\dot{\rho}_{\rm I} & = & - {\textstyle {{\rm i} \over 2}} \Omega \, \left[ {\rm e}^{{\rm i} \varphi}  \, c + {\rm e}^{- {\rm i} \varphi}   \, c^\dagger , \rho_{\rm I}\right] \nonumber \\
&& + {\textstyle {1 \over 2}}  \kappa \left( 2 c \rho_{\rm I} c^{\dagger} - \left[ c^{\dagger} c , \rho_{\rm I} \right]_+ \right) \, , 
\end{eqnarray}
where $\kappa$ denotes the spontaneous decay rate of the cavity. Now suppose a detector with efficiency $\eta$ monitors the spontaneous leakage of photons and a feedback loop is activated and applies a unitary operator $R$ to the resonator field whenever a photon is detected. Proceeding as suggested in the previous section, we find that the master equation of the cavity equals  
\begin{eqnarray} \label{Cavity master equation feedback0}
\dot{\rho}_{\rm I} & = & - {\textstyle {{\rm i} \over 2}} \Omega \left[ {\rm e}^{{\rm i} \varphi} \, c + {\rm e}^{- {\rm i} \varphi} \, c^\dagger , \rho_{\rm I}\right] \nonumber \\
&& + \eta \cdot {\textstyle {1 \over 2}} \kappa \left( 2 R c \rho_{\rm I} c^{\dagger} R^{\dagger} - \left[ c^{\dagger} c , \rho_{\rm I} \right]_+ \right) \nonumber \\
&& + \left(1 - \eta \right) \cdot {\textstyle {1 \over 2}} \kappa \left( 2 c \rho_{\rm I} c^{\dagger} - \left[ c^{\dagger} c , \rho_{\rm I} \right]_+ \right)
\end{eqnarray}
in this case. The second line in this equation takes the effect of the feedback loop into account, while the third line corresponds to undetected photon emission events. Simplifying Eq.~(\ref{Cavity master equation feedback0}) yields
\begin{eqnarray} \label{Cavity master equation feedback}
\dot{\rho}_{\rm I} & = & - {\textstyle {{\rm i} \over 2}} \Omega \left[ {\rm e}^{{\rm i} \varphi} \, c + {\rm e}^{- {\rm i} \varphi} \, c^\dagger , \rho_{\rm I}\right] \nonumber \\
&& + \eta \kappa \left( R c \rho_{\rm I} c^{\dagger} R^{\dagger} - c \rho_{\rm I} c^{\dagger} \right) \nonumber \\
&& + {\textstyle {1 \over 2}} \kappa \left( 2 c \rho_{\rm I} c^{\dagger} - \left[ c^{\dagger} c , \rho_{\rm I} \right]_+ \right) \, . 
\end{eqnarray}
In the following, we consider instantaneous feedback in the form of a very short strong laser pulse, meaning that $R$ can be written as 
\begin{eqnarray} \label{displacement}
R & = & D(\beta) 
\end{eqnarray}
with $D(\beta)$ being a displacement operator of the form
\begin{eqnarray} \label{displacement2}
D(\beta) &=& \exp \left( \beta \, c^\dagger - \beta^* \, c \right) \, .
\end{eqnarray}
Here $\beta$ is a complex number, which characterises the strength of the feedback pulse. Without loss of generality we may take $\beta$ to have any phase we want by absorbing any unwanted phase factor into the definition of the cavity photon annihilation operator $c$.

\subsection{Unravelling into quantum trajectories} \label{lastnumber}

We now have all the information needed to analyse the time evolution of the electromagnetic field inside the resonator under the condition of no photon emission and in the case of the detection of a photon. In this subsection, we introduce the equations needed to numerically simulate all possible quantum trajectories of the experimental setup in Fig.~\ref{Cavity-Prep}. As we shall see below, the cavity field remains always in a coherent state. 

\subsubsection{No photon time evolution}

Substituting Eq.~(\ref{Cavity interaction Hamiltonian}) into Eq.~(\ref{master2}) we obtain an equation of the same form as Eq.~(\ref{master22}). Subsequently comparing Eq.~(\ref{master2}) and (\ref{master22}), we find that  
\begin{eqnarray} \label{H-cond}
H_{\rm cond} &=& {\textstyle {1 \over 2}} \hbar \Omega \left( {\rm e}^{{\rm i} \varphi} c + {\rm e}^{-{\rm i} \varphi} c^{\dagger} \right) - {\textstyle {{\rm i} \over 2}} \hbar \kappa \, c^{\dagger} c \, . 
\end{eqnarray}
This conditional Hamiltonian describes the dynamics of the cavity field under the condition of no photon emission. The corresponding conditional time evolution operator,
\begin{eqnarray}
U_{\rm cond} \left(t+ \Delta t , t\right) &=& \exp \left( - {{\rm i} \over \hbar} H_{\rm cond} \Delta t \right) 
\end{eqnarray}
is given by
\begin{eqnarray} \label{BCH U}
U_{\rm cond} \left(t + \Delta t , t \right) & = & {\rm exp} \left[- {\textstyle {{\rm i} \over 2}} \Omega \left({\rm e}^{{\rm i} \varphi} c + {\rm e}^{-{\rm i} \varphi} c^{\dagger} \right) \Delta t \right] \nonumber \\
&& \times {\rm exp} \left[- {\textstyle {1 \over 2}} \kappa \, c^{\dagger} c \, \Delta t \right] 
\end{eqnarray}
up to terms of order $(\Delta t )^2$. Calculating the effect of the second exponential in this equation onto a coherent state $|\alpha (t) \rangle$ is best done using the Fock basis. Moreover using the general properties of displacement operators to evaluate the first exponential in Eq.~(\ref{BCH U}), we eventually see that 
\begin{eqnarray} \label{exact}
|\alpha (t+ \Delta t) \rangle &=& U_{\rm cond} \left(t+\Delta t , t\right) \, |\alpha (t) \rangle / \| \cdot \| \nonumber \\
&=& \left| {\rm e}^{- {1 \over 2} \kappa \Delta t} \, \alpha(t) + {\textstyle {1 \over 2}} \Omega  {\rm e}^{- {\rm i} \varphi} \, \Delta t \right \rangle
\end{eqnarray}
is the normalised state of the cavity field under the condition of no photon emission in $(t,t+\Delta t)$. This equation tells us that
\begin{eqnarray}
\dot \alpha(t) &=& - {\textstyle {1 \over 2}} \kappa \, \alpha(t) + {\textstyle {1 \over 2}} \Omega  {\rm e}^{- {\rm i} \varphi} 
\end{eqnarray}
without any approximations. Solving this differential equation for an initial coherent state  $|\alpha (0) \rangle$ shows that
\begin{eqnarray} \label{State evolution}
\alpha(t) &=& {\rm e}^{- {1 \over 2} \kappa t} \, \alpha(0) + {\Omega \over \kappa} \left( 1 - {\rm e}^{- {1 \over 2} \kappa t} \right) \, {\rm e}^{- {\rm i} \varphi} 
\end{eqnarray}
under the condition of no photon emission in $(t,t+\Delta t)$. If no photon is emitted for a relatively long time $t \gg 1/\kappa$, then the state of the resonator becomes
\begin{eqnarray} \label{alphass}
|\alpha_{\rm ss} \rangle &=& \left | {\Omega \over \kappa} \, {\rm e}^{- {\rm i} \varphi} \right \rangle \, .
\end{eqnarray}
This state is invariant under the no-photon time evolution of the system. Using Eq.~(\ref{master3333}), the calculations which lead to Eq.~(\ref{exact}) moreover reveal that 
\begin{eqnarray} \label{P0}
P_0 (\Delta t) &=& \exp \left[ - |\alpha(t)|^2 \left( 1 - {\rm e}^{- \kappa \Delta t} \right) \right]
\end{eqnarray}
is the probability for no photon emission in a short time interval $(t, t+\Delta t)$.

\subsubsection{Spontaneous photon emission}

To determine the density matrix $\rho_{\rm S}^{\neq} (t)$, which describes the cavity field immediately after a photon emission, we now substitute the system-bath Hamiltonian $H_{\rm SB \, I}$ in Eq.~(\ref{Cavity interaction Hamiltonian}) into Eq.~(\ref{master38}). Evaluating all integrals, we find that 
\begin{eqnarray}
\dot{\rho}_{\rm S}^{\neq} (t) &=& \kappa \, c \, \rho_{\rm S} (t) \, c^\dagger 
\end{eqnarray}
within the usual standard approximations. Since the coherent states are eigenstates of the photon annihilation operator $c$, the emission of a photon does not change the state of the cavity and
\begin{eqnarray} \label{more}
|\alpha(t + \Delta t) \rangle &=& |\alpha(t) \rangle \, , 
\end{eqnarray}
if the resonator is initially in a coherent state and no feedback pulse is applied. If the photon emission triggers a feedback pulse, then \begin{eqnarray} \label{41} 
|\alpha(t + \Delta t) \rangle &=& D(\beta) \, |\alpha(t) \rangle = |\alpha(t) + \beta \rangle \, .
\end{eqnarray}
In the next section, we use this equation as well as Eqs.~(\ref{State evolution}), (\ref{P0}) and (\ref{more}) to numerically generate the possible quantum trajectories of the experimental setup in Fig.~\ref{Cavity-Prep}. In every time step of the simulation, we test for a photon emission. A further test is performed to decide, if feedback is applied or not, while taking into account the likeliness for such an event to occur. 

\subsection{Long term behaviour}

\begin{figure*}[t]
\centering
\includegraphics[width=\textwidth]{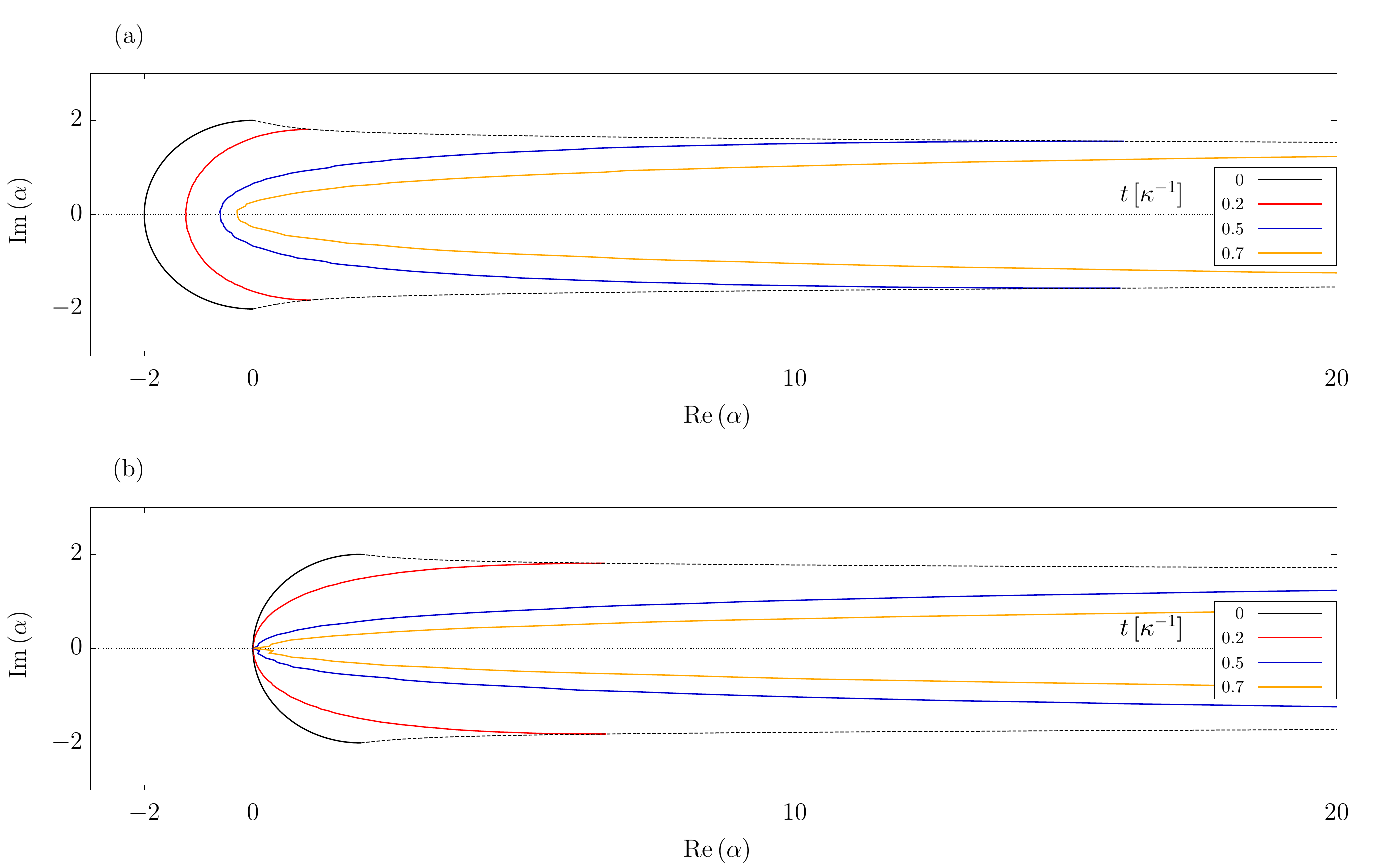} \\
\caption{[Color online] (a) Phase diagram illustrating the dynamics of the single-mode coherent states $|\alpha \rangle$ of the cavity field during the measurement stage. The initial states of the resonator form a circle centred about the origin. The lines show the occupied state space of the states corresponding to $\varphi \in [{\pi \over 2},{3 \pi \over 2}]$ at a later time $t$. As time elapses, the circle turns into an increasingly stretched ellipse. States that correspond to different phases $\varphi$ move further and further away from each other. (b) Dynamics of the cavity field under the condition of a photon emission at $t=0$, which triggered an instantaneous feedback pulse. Both graphs are the result of a quantum jump simulation based on the calculations in Section \ref{Cavity}, where we assume a detector efficiency of $\eta = 0.5$ and consider $10^6$ repetitions of the experiment. Here the feedback pulse is given by $\beta = |\alpha|$ with $\alpha = 2$.  The dashed lines represent the trend of the evolution of the states corresponding to $\varphi = {\pi \over 2}$ and $\varphi = {3 \pi \over 2}$.} \label{Phase-space}
\end{figure*}

Finally, we have a closer look at the stationary states of the master equations (\ref{Cavity master equation}) and (\ref{Cavity master equation feedback}) of a laser-driven cavity with and without instantaneous feedback.

\subsubsection{Convergence without feedback}

In the previous subsection, we have seen that the coherent state $|\alpha_{\rm ss} \rangle$ in Eq.~(\ref{alphass}) is invariant under the no-photon time evolution of a laser-driven optical cavity. Eq.~(\ref{more}) shows that this state is also invariant under the emission of a photon. Consequently, $|\alpha_{\rm ss} \rangle$ is the stationary state of a laser-driven optical cavity without feedback. Once the cavity reaches this state, it no longer evolves in time. Indeed, one can easily check that the corresponding density matrix 
$\rho_{\rm ss} = |\alpha_{\rm ss} \rangle \langle \alpha_{\rm ss} |$ solves $\dot \rho =0$.

\subsubsection{Divergence with feedback}

Combining the stationary state condition $\dot \rho =0$ with the master equation in Eq.~(\ref{Cavity master equation feedback}), we now calculate the stationary state of the laser-driven optical cavity inside an instantaneous feedback loop. From the discussion in the previous subsection, we know that the field inside the resonator in this case too remains always in a coherent state, if initially coherent. This implies that the stationary state, if it is ever reached, has to be of the form
\begin{eqnarray}
\rho_{\rm ss} &=& \int_{C \!\!\!\! I} {\rm d} \alpha \, P(\alpha) \,|\alpha \rangle \langle \alpha | \, ,
\end{eqnarray}
i.e.~a statistical mixture of coherent states $|\alpha \rangle$ with weighting $P(\alpha)$. However, the master equation (\ref{Cavity master equation feedback}) does not possess a stationary state of this form. From this we conclude that the laser-driven cavity with instantaneous feedback that we consider in this paper never reaches a stationary state \cite{Masalov,students}. It exhibits a much richer dynamics than what was previously assumed \cite{Kilin,Golub}. This even applies if the continuous laser driving is turned off, unless the cavity is initially empty.

Fig.~\ref{Phase-space} illustrates the non-linear dynamics of an optical cavity inside an instantaneous quantum feedback loop with the help of a so-called phase diagram. This diagram represents coherent states $|\alpha \rangle$ as points by using the real part and the imaginary part of $\alpha$ as coordinates. It is the result of a numerical simulation which averages $\alpha (t)$ over a large number of quantum trajectories. Different times $t$ and a wide range of initial states $|\alpha \rangle$ with $\varphi \in [{\pi \over 2},{3\pi \over 2}]$ and with $\alpha$ as in Eq.~(\ref{polar}) are considered. As one can see, the half circle representing these initial states deforms rapidly into an increasingly stretched ellipse, thereby constantly increasing the phase space volume occupied by the cavity field. In Fig.~\ref{Phase-space}(b), the cavity field is initially in the same state as in Fig.~\ref{Phase-space}(a) but experiences a feedback pulse at $t=0$ with $\beta = |\alpha|$. In this case, the constant growth and stretching of the phase space volume of the cavity field is even more pronounced. For example, a cavity in an initial coherent state with $\varphi = \pi$ and a photon detection at $t=0$ never emits another photon and never experiences another feedback pulse. On the contrary, a cavity with $\varphi \neq \pi$ is likely to emit many photons, thereby attracting an exponentially-increasing number of feedback pulses. As a result, the distance between two coherent states $|\alpha_1(t) \rangle$ and $|\alpha_2(t) \rangle$ corresponding to two different phases $\varphi_1$ and $\varphi_2$ increases rapidly in time. 

\section{Quantum-enhanced metrology} \label{Metrology}

Now we have all the tools needed to analyse the quantum metrology scheme illustrated in Fig.~\ref{Cavity-Prep}. It consists of two main stages: 
\begin{enumerate}
\item {\bf The preparation stage}. A continuous laser field experiences an unknown phase shift $\varphi$ before entering an optical cavity, as illustrated in Fig.~\ref{Cavity-Prep}(a). The main purpose of this stage is to prepare the field inside the resonator in a coherent state, which depends on $\varphi$. For simplicity, we assume that the cavity is driven for a time which is relatively long compared to the time scale given by the laser Rabi frequency and the cavity decay rate. This approach prepares the resonator in its stationary coherent state $|\alpha_{\rm ss} \rangle$ in Eq.~(\ref{alphass}) with the phase $\varphi$ encoded into the phase of $\alpha^{\rm ss}$.
\item {\bf The measurement stage}. Here the continuous laser driving is turned off. Instead the optical cavity evolves freely, while experiencing instantaneous feedback pulses, as illustrated in Fig.~\ref{Cavity-Prep}(b). These are triggered by the observation of a spontaneously emitted photon with a finite detector efficiency $\eta$. We assume that every feedback pulse displaces the field inside the cavity by an amount $\beta$ given by
\begin{eqnarray} \label{beta}
\beta &=& - {\rm i} \, |\alpha_{\rm ss}| 
\end{eqnarray}
which is independent of $\varphi$. This means, the feedback laser provides a reference frame. The measured phase is indeed the relative phase between the phase of the driving laser used during the preparation stage and the phase of the feedback laser (with respect to the interaction picture). All photon detection times should be registered.
\end{enumerate} 
As we shall see below, temporal quantum correlations reveal information about $\varphi$ with an accuracy above the standard quantum limit. 

\subsection{Resource counting}

\begin{figure}[t]
\centering
\includegraphics[width=0.48\textwidth]{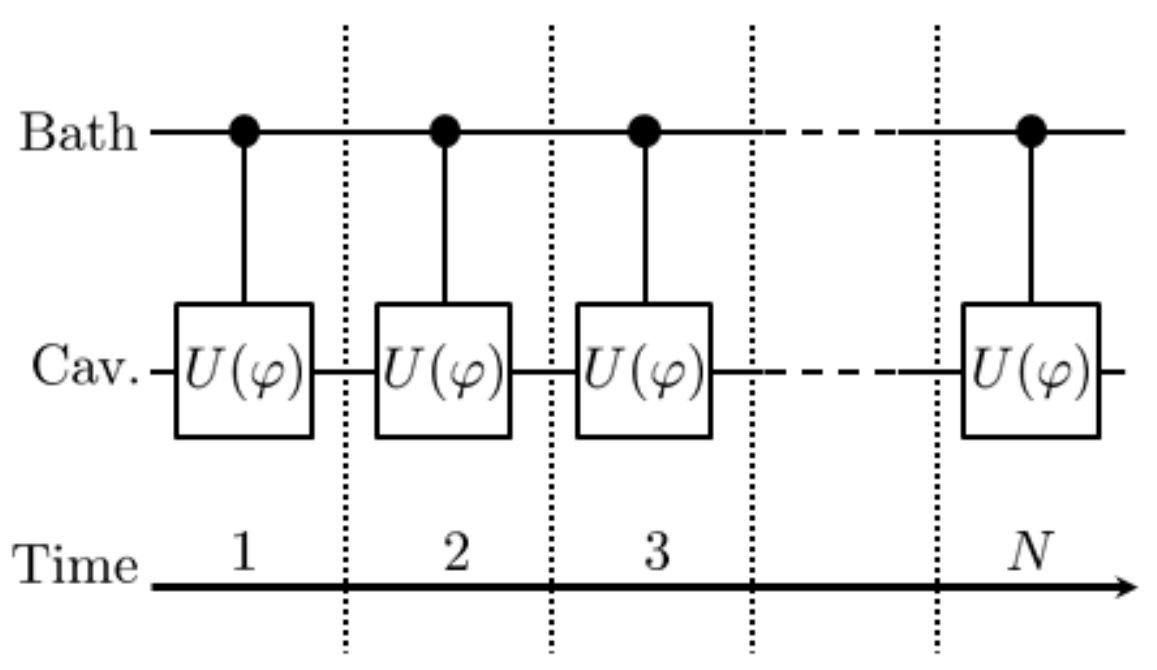}
\caption{Circuit diagram of the time evolution of the experimental setup in Fig.~\ref{Cavity-Prep} during the measurement stage. The black dots indicate that the bath is measured in every time step, $n=1,\dots ,N$, and when a photon emission is detected triggers the operator $U(\varphi)$ to act on the cavity. This process provides information about the state of the cavity and the unknown phase, $\varphi$.} \label{Circuit}
\end{figure}

To identify the main resource, $N$, of our metrology scheme, we adopt the same approach as Zwierz {\em et al.}~\cite{Kok} and assume that $N$ equals the query complexity of our scheme. Each time the phase $\varphi$ is probed, a resource is used. In every time step, we perform a (conditional) phase dependent operation on the system. Continuously observing the leakage of photons through the cavity mirrors means a continuous probing of the unknown phase $\varphi$. As illustrated in Fig.~\ref{Circuit}, every time step can be seen as one query posed and hence provides one resource count. The amount of time $T$, which the system spends within the measurement stage during each repetition of the experiment, is therefore the most relevant resource of our quantum metrology scheme. To calculate $\Delta \varphi$ as a function of $T$, we now simulate a relatively large number of quantum trajectories of the experimental setup in Fig.~\ref{Cavity-Prep} using the methodology which we introduced in the previous section and then use the error propagation formula in Eq.~(\ref{Error}) to analyse the precision of the proposed experiment. For completeness and to allow for a comparison with other quantum metrology schemes, we also consider the mean number of photons passing through the unknown phase $\varphi$ as a resource $N$. In our scheme, this number is essentially given by the mean number of photons $|\alpha_{\rm ss}|^2$ inside the resonator at the end of the preparation stage. 

\subsection{Accuracy of intensity measurements} \label{Results1}

\begin{figure}[t]
\centering
\includegraphics[width=0.48\textwidth]{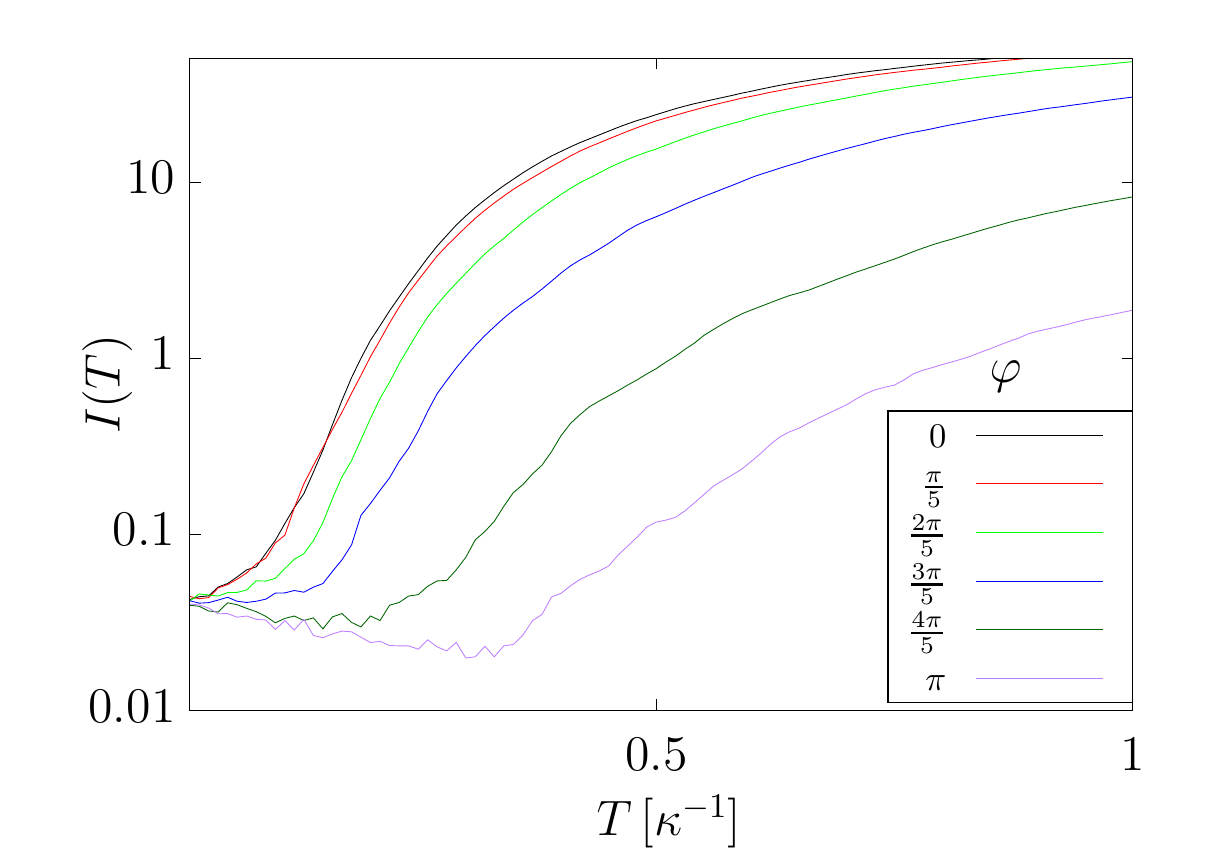}
\caption{[Color online] Average intensity, $I(T)$, as a function of the time, $T$, after the preparation of the initial coherent state, $|\alpha_{\rm ss} \rangle$, for various unknown phases, $\varphi$. This simulation assumes $| \alpha_{\rm ss} |^2 = 4$, $\eta = 0.5$ and averages over $10^6$ trajectories.}
\label{int-time}
\end{figure}

\begin{figure}[t]
\centering
\includegraphics[width=0.48\textwidth]{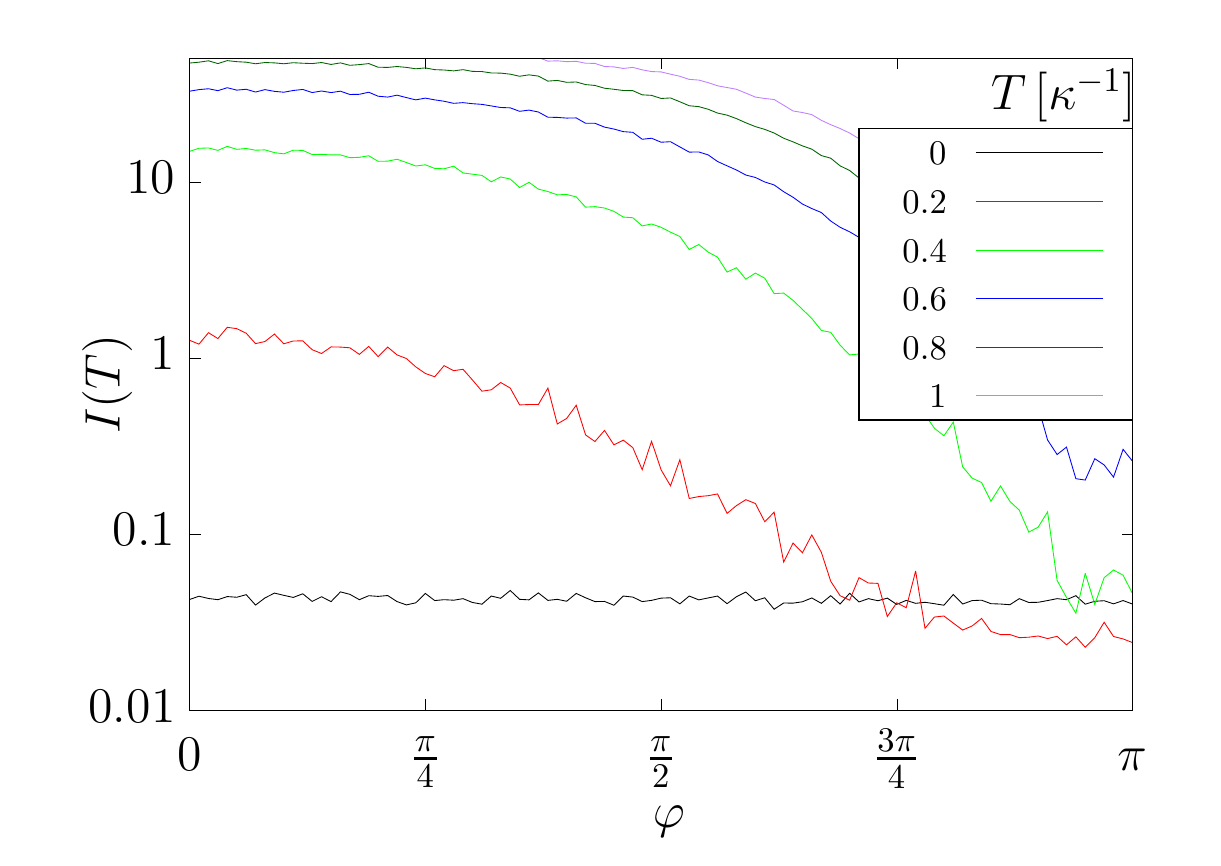}
\caption{[Color online] Average intensity, $I(T)$, as a function of the unknown phase, $\varphi$, for different times, $T$. As in Fig.~\ref{int-time}, we have $| \alpha_{\rm ss} |^2 = 4$, $\eta = 0.5$ and average over $10^6$ trajectories.}
\label{int-phase}
\end{figure}

Let us first have a closer look at the average photon emission rate $I(T)$ of the cavity at a time $T$ after the preparation of the initial coherent state $|\alpha_{\rm ss} \rangle$ in Eq.~(\ref{alphass}), which depends on the unknown phase $\varphi$. To calculate $I(T)$ numerically, we divide the time interval $[0,T]$ into relatively short time intervals $\Delta t$.  We then use the quantum jump approach \cite{Reset,Carmichael,Molmer} to simulate a relatively large number of possible quantum trajectories of the cavity and average over the respective number of photon emissions in $(T, T + \Delta t)$. The result of this simulation is shown in Figs.~\ref{int-time} and \ref{int-phase}. While Fig.~\ref{int-time} shows the average photon emission rate $I(T)$ as a function of $T$ for different phases $\varphi$, Fig.~\ref{int-phase} shows the $I(T)$ as a function $\varphi$ for different times $T$. Both logarithmic plots illustrate that the dynamics of the mean number of photons inside the resonator depends indeed very strongly on the initial coherent state $|\alpha_{\rm ss} \rangle$ of the cavity field.  

\begin{figure}[t]
\centering
\includegraphics[width=0.48\textwidth]{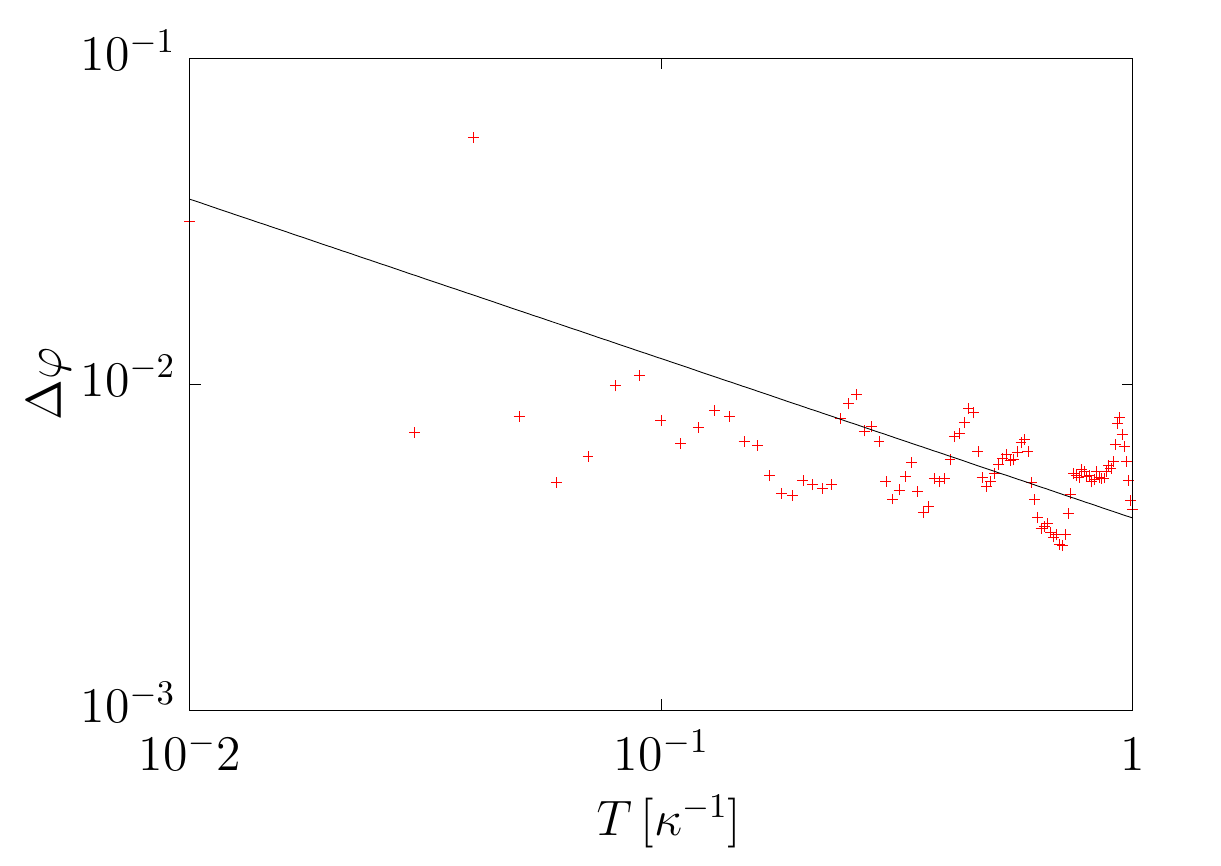}
\caption{[Color online] Dependence of the accuracy, $\Delta \varphi$, on the length of the measurements stage, $T$, in the case of intensity measurements. Here, $\varphi = 0.3 \, \pi$, $| \alpha_{\rm ss} |^2 = 4$, $\eta = 0.5$ and we averaged over $10^6$ trajectories. The black line illustrates the approximate fit given in Eq.~(\ref{Limitsclass2}).}
\label{int-var}
\end{figure}

Next we investigate the accuracy of a measurement, which uses the strong dependence of $I(T)$ on $\varphi $ to deduce information about $\varphi$. Figs.~\ref{int-time} and \ref{int-phase} show that this dependence is maximised for $\varphi $ around $0.3 \, \pi$. We therefore consider in the following the signal $M = I(T)$ and $\varphi = 0.3 \, \pi$ as an example and calculate the accuracy of the proposed quantum metrology scheme $\Delta \varphi$ using the error propagation formula in Eq.~(\ref{Error}) as a function of $T$. The variance in this equation is obtained through statistical analysis of the simulation data, while the sensitivity is found by finding the gradient between two very close phases. Again we average over a large number of quantum trajectories. The result of this numerical simulation is shown in Fig.~\ref{int-var}. To a very good approximation, we find that
\begin{eqnarray} \label{Limitsclass2}
\Delta \varphi (T) &\propto & T^{-0.49} ~~ {\rm for} ~~\varphi = 0.3 \, \pi \, .
\end{eqnarray}
This means, using only intensity measurements, the experimental setup in Fig.~\ref{Cavity-Prep} does not allow us to beat the standard quantum limit in Eq.~(\ref{Limitsclass}). In fact, we almost saturate this limit. This is to be expected as the dynamics of the mean number of photons inside the cavity obeys a master equation, which represents a set of {\em linear} differential equations. Moreover, intensity measurements are essentially classical measurements. Finally, it should be noted that the fit In Eq.~(\ref{Limitsclass2}) is an approximation, as the data appears to be showing a more complex dependence. However, it seems safe to conclude that the scaling of $\Delta \varphi (T) $ does not exceed the standard quantum limit.

\subsection{Accuracy of second-order correlation function measurements} \label{Results2}

A comparison between Figs.~\ref{Phase-space}(a) and (b) suggests that measurements of the joint probability to detect a photon at a time $t$ {\em and} at a time $t'$ should be able to reveal information about $\varphi$ more efficiently than measurements of the average photon intensity $I(T)$. This joint probability is known to quantum opticians as the second-order photon correlation function $G^{(2)}(t,t')$. Hence, according to probability theory, $G^{(2)}(t,t')$ equals
\begin{eqnarray} \label{G2}
G^{(2)}(t,t') &\equiv & I(t|t') I(t') \, ,
\end{eqnarray}
where $I(t|t')$ denotes the probability for the detection of a photon at a time $t$ conditional on the detection of a photon at $t'$.  
Second-order correlation functions are usually normalised by the product of the photon emission rate at $t'$ and at $t$. Taking this into account and dividing Eq.~(\ref{G2}) by $I(t') I(t)$, we define the renormalised second-order correlation function, $g^{(2)}(t,t')$, by
\begin{eqnarray}
g^{(2)}(t,t') &\equiv & {I(t|t') \over I(t)} \, .
\end{eqnarray}
This correlation function describes correlations between photon emission events without depending on the detector efficiency $\eta$ with which these events are registered. It can therefore be measured accurately, even when using imperfect detectors with $\eta < 1$. 

\begin{figure}[t]
\centering
\includegraphics[width=0.48\textwidth]{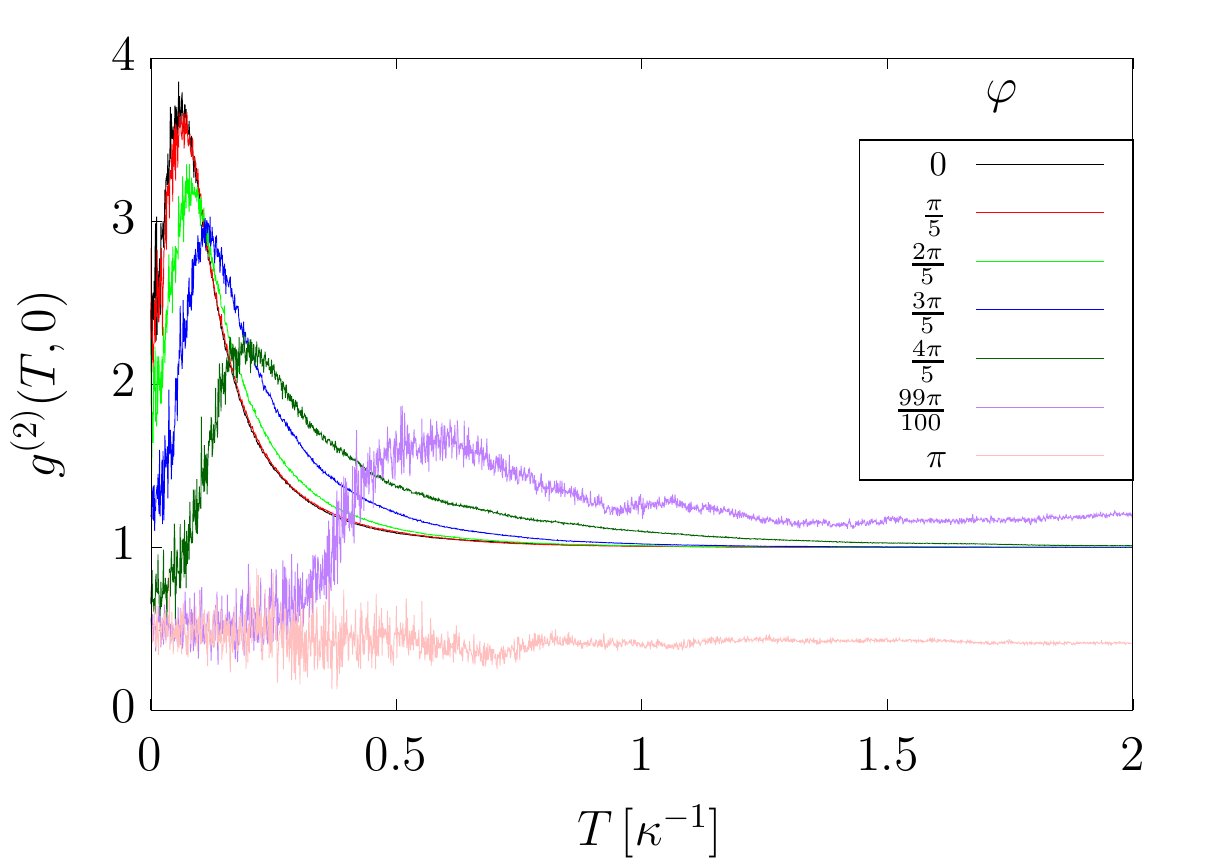}
\caption{[Color online] Second-order correlation function $g^{(2)}(T,0)$, as a function of the duration of the measurement stage, $T$, for various phases $\varphi$. Again we assume $| \alpha_{\rm ss} |^2 = 4$, $\eta = 0.5$ and averages over $10^6$ trajectories.}
\label{corr-time}
\end{figure}

\begin{figure}[t]
\centering
\includegraphics[width=0.48\textwidth]{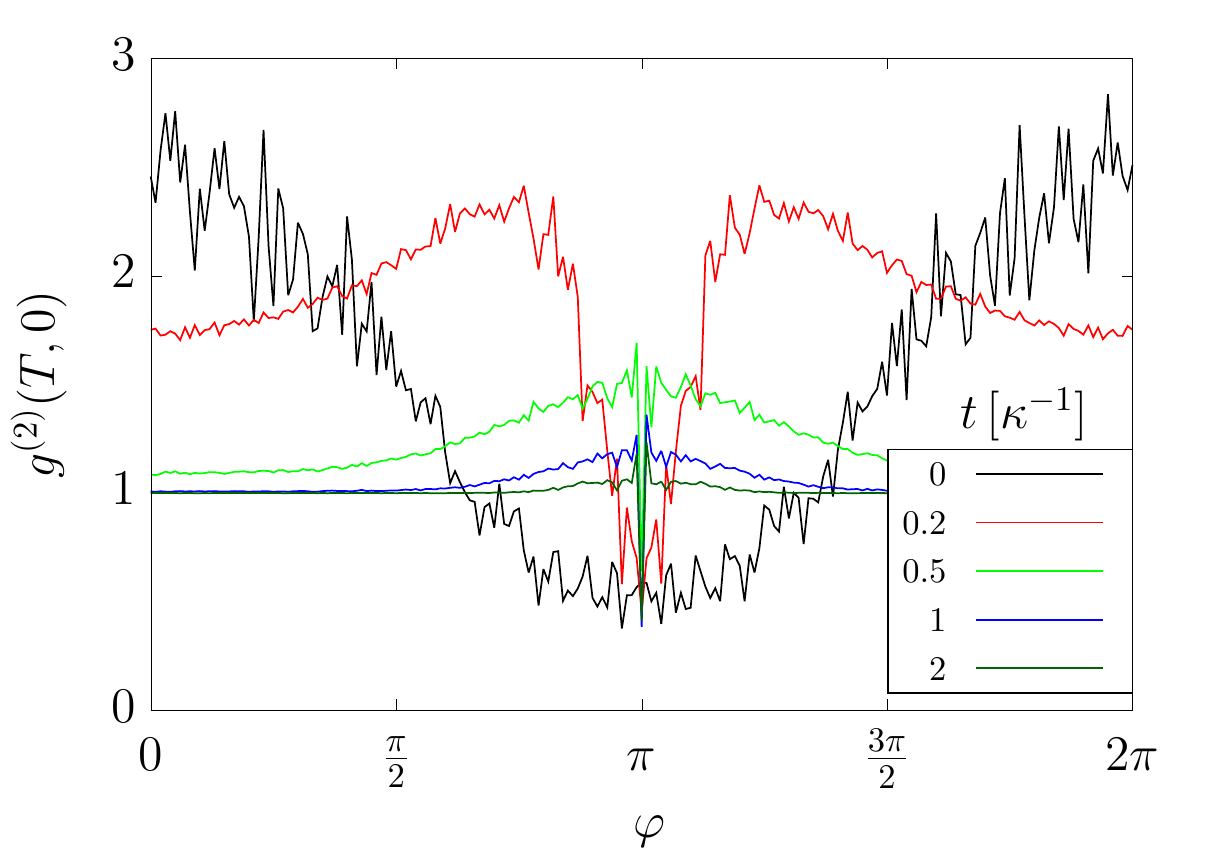}
\caption{[Color online] Second-order correlation function $g^{(2)}(T,0)$ as a function of the unknown phase, $\varphi$ for various times, $T$, with $|\alpha_{\rm ss} |^2 = 4$ and $\eta = 0.5$ averaged over $10^6$ trajectories.}
\label{corr-phase}
\end{figure}

\begin{figure}[t]
\centering
\includegraphics[width=0.48\textwidth]{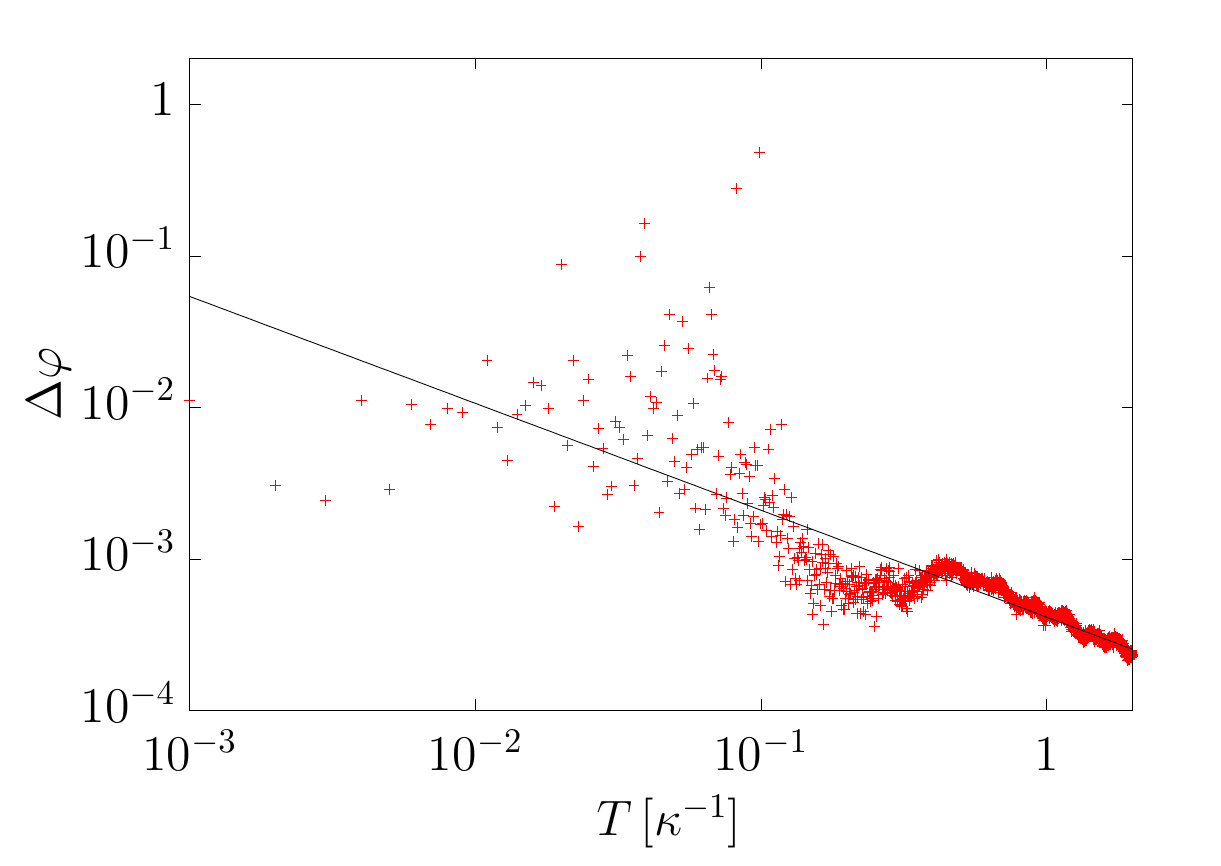}
\caption{[Color online] Accuracy $\Delta \varphi$ of the proposed metrology scheme as a function of the duration of the measurement stage, $T$, for measurements of the second-order correlation function $g^{(2)}(T,0)$ around $\varphi = \pi$ to maximise the sensitivity of the proposed scheme. As before, we assume $|\alpha_{\rm ss}|^2 = 4$, $\eta = 0.5$ and average over $10^6$ trajectories. The black line shows the approximate solution in Eq.~(\ref{Limitsclass4}).}
\label{corr-var}
\end{figure}

In the following, we assume that $M=g^{(2)}(T,0)$ is the actual measurement signal used to obtain information about the unknown phase $\varphi$. To determine the accuracy $\Delta \varphi$ of this approach as a function of the length $T$ of the measurement stage, we again simulate a relatively large number of quantum trajectories and average over all of them. The results of this simulation are shown in Figs.~\ref{corr-time} and \ref{corr-phase}, which are analogous to Figs.~\ref{int-time} and \ref{int-phase} in the previous subsection. As expected, the correlation function $g^{(2)}(T,0)$ too exhibits a very strong $\varphi$-dependence. This dependence is most pronounced when $\varphi = \pi $. As suggested by Fig.~\ref{Phase-space}, we have $g^{(2)}(T,0) = 0$ for sufficiently large detector efficiencies $\eta$ and $\varphi = \pi $, while $g^{(2)}(T,0)$ rapidly tends to unity for all other angles. Indeed Fig.~\ref{corr-time} shows very large differences between neighbouring curves, when $\varphi$ is close to $\pi$, even when $\varphi$ is varied only by a relatively small amount. Moreover Fig.~\ref{corr-phase} shows a distinct spike at $\varphi = \pi $ as a function of $\varphi$. This spike in the second-order correlation function is what allows us to distinguish this phase with a very high accuracy $\Delta \varphi$ from other close-by values of $\varphi$ due to a very high visibility $|\partial M / \partial \varphi |$.

\begin{figure}[t]
\centering
\includegraphics[width=0.48\textwidth]{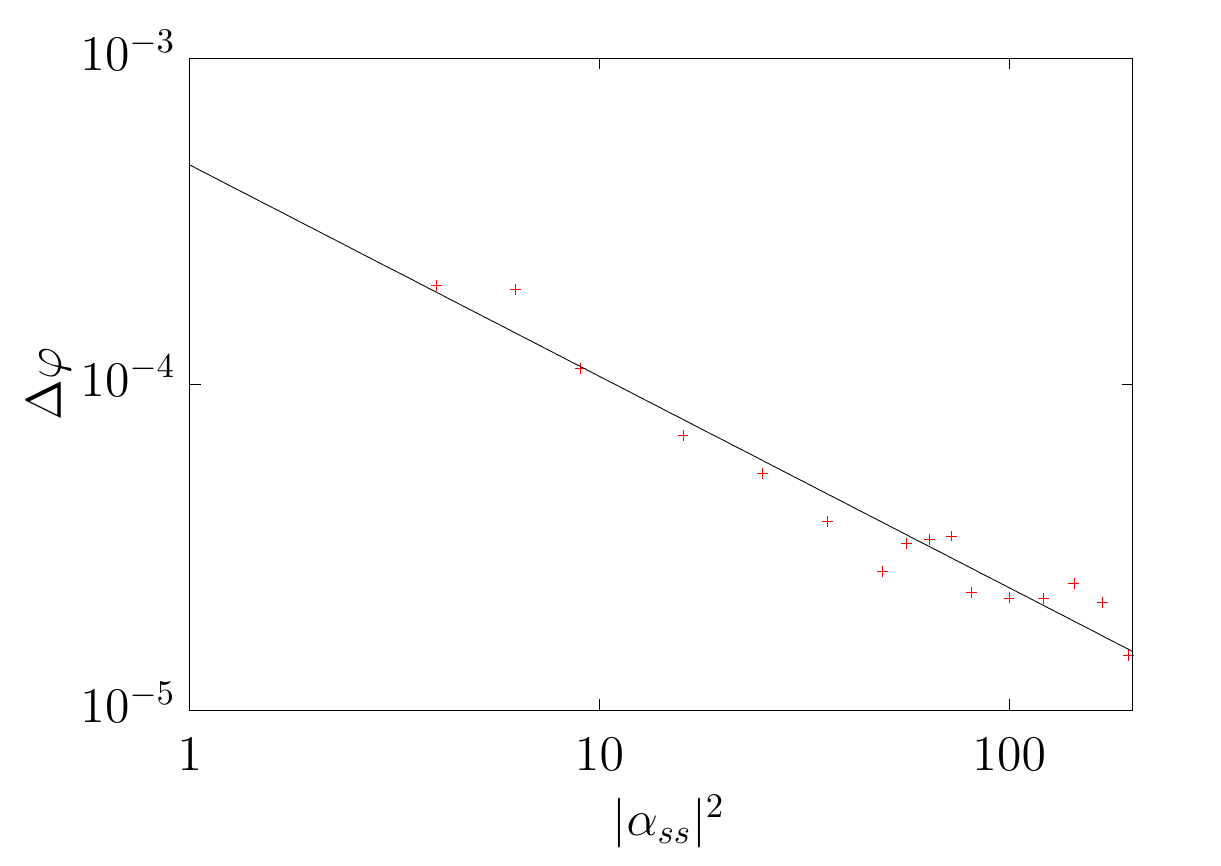}
\caption{[Color online] Accuracy, $\Delta \varphi$, of the proposed metrology scheme as a function of the initial mean photon number, $|\alpha_{\rm ss}|^2$, for measurements of the second-order correlation function $g^{(2)}(T,0)$ and $\varphi = \pi$. Here, $\eta = 0.5$ and we average over $10^6$ trajectories. To remove noisy fluctuations in the signal in time, we take a sample of uncertainties over a fixed period of time, find the average uncertainty in that period and compare this average to the same time average for other initial states. The black line shows the approximate solution in Eq.~(\ref{Limitsclass5}).}
\label{corr-var-photon}
\end{figure}

This is confirmed by Fig.~\ref{corr-var} which shows the dependence of $\Delta \varphi$ on the resource $T$ for phase measurements based on the second-order correlation function for the optimal case of $\varphi = \pi$. To calculate this quantity we use again the error propagation formula in Eq.~(\ref{Error}) and average over a relatively large number of quantum trajectories. We now find that
\begin{eqnarray} \label{Limitsclass4}
\Delta \varphi (T) &\propto & T^{-0.71} ~~ {\rm for} ~~ \varphi = \pi 
\end{eqnarray}
to a very good approximation. This accuracy clearly beats the standard quantum limit. In other words, measurements of the second-order photon correlation function of the photon statistics of an optical cavity inside an instantaneous quantum feedback loop can be very sensitive to phase fluctuations. 

This is not surprising, since measurements of the second-order photon correlation function $g^{(2)}(T,0)$ require the detection of single photons. This is different from intensity measurements which are essentially classic measurements. These can be done without high-resolution single-photon detection. Moreover, second-order photon correlations are an intrinsic property of the individual quantum trajectories of the cavity field. They cannot be calculated with the help of a linear master equations but require the quantum jump approach \cite{Reset,Molmer,Carmichael}, which we introduced in Section \ref{lastnumber}. The conditional dynamics of the individual trajectories of the cavity field is in general non-linear. For example, between photon emission, the cavity field evolves in a non-linear fashion with the non-Hermitian conditional Hamiltonian $H_{\rm cond}$ in Eq.~(\ref{H-cond}), which requires a constant renormalisation of the state vector of the quantum system.  In summary, it is the measurement of the temporal quantum correlations in an open quantum system that allows us to exceed the standard quantum limit. This observation is consistent with analogous observations by other authors \cite{Kok2,Kok3,Ref1,Ref2,Ref3,Braun,Openmet}.

\begin{figure}[t]
\centering
\includegraphics[width=0.48\textwidth]{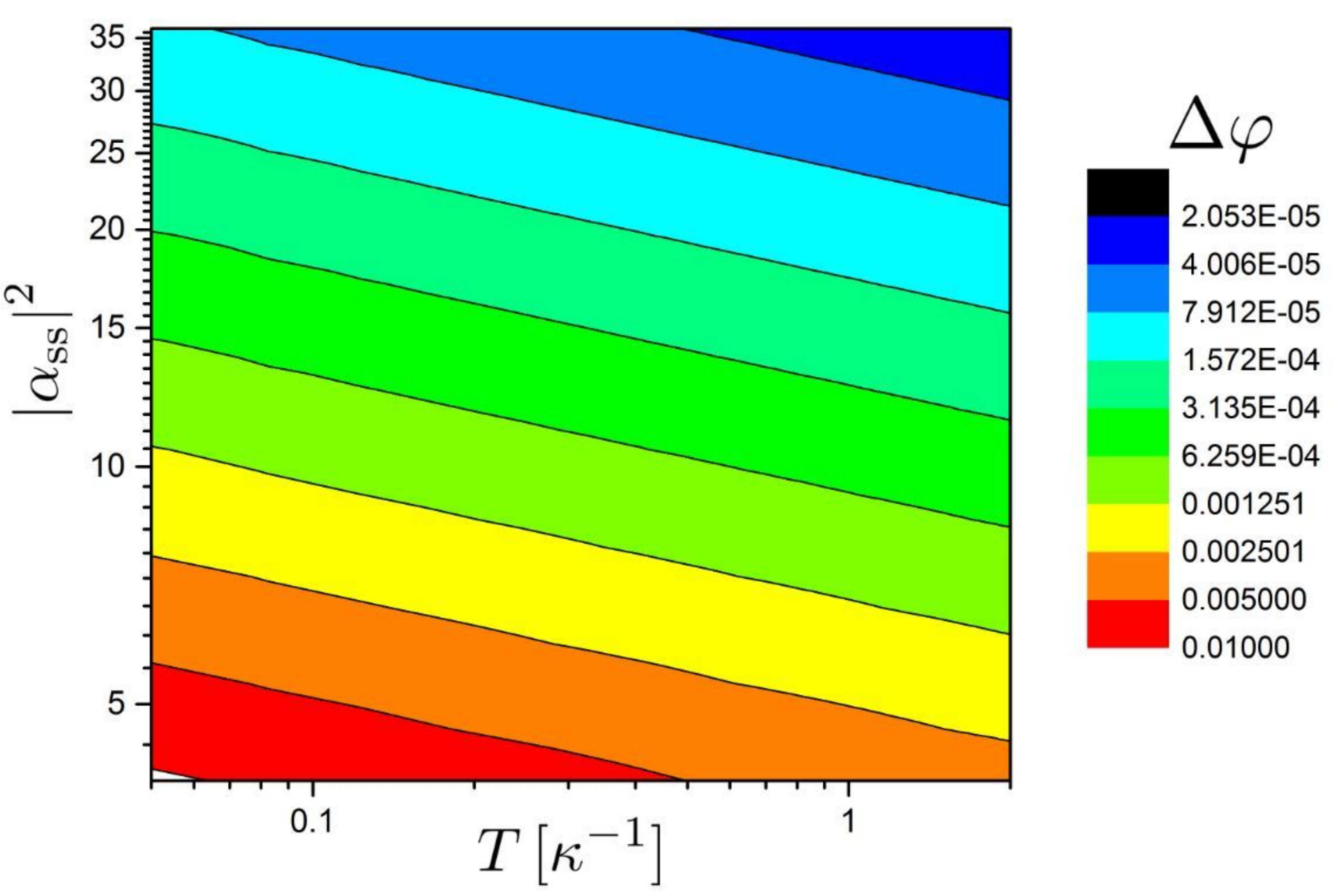}
\caption{[Color online] Log-Log plot of the accuracy, $\Delta \varphi$, when both the duration of the measurement stage, $T$, and the initial mean number of photons, $| \alpha_{\rm ss} |^2$, is taken into account and the second-order correlation function $g^{(2)}(T,0)$ is analysed. Here we have $\eta = 0.5$ and we consider $10^6$ repetitions of the experiment.}
\label{multi-var}
\end{figure}

A more standard method of resource counting in quantum metrology is to consider the average number of photons that passed through the unknown phase $\varphi$ as the resource $N$. This approach can also be applied to the quantum-enhanced metrology scheme which we propose here. Performing quantum jump simulations, averaging over many quantum trajectories and using again the error propagation formula in Eq.~(\ref{Error}) with $M= g^{(2)}(T,0) $, we now calculate the dependence of $\Delta \varphi$ on the average population of the initial coherent state inside the cavity, which is given by $\left| \alpha_{\rm ss} \right|^2$. The result is shown in Fig.~\ref{corr-var-photon}. For the parameters that we consider here, we find that
\begin{eqnarray} \label{Limitsclass5}
\Delta \varphi (\left| \alpha_{\rm ss} \right|^2) &\propto & \left( \left| \alpha_{\rm ss} \right|^2 \right)^{-0.65}  ~~ {\rm for} ~~ \varphi = \pi
\end{eqnarray}
to a very good approximation. Eq.~(\ref{Limitsclass5}) too clearly beats the standard quantum limit. In practical applications, it might be best to consider both the duration of the measurement stage and the number of photons that passed through the sample as a resource. Numerical results for such an experiment are shown in Fig.~\ref{multi-var}. When we have two scalable resources, our scheme allows more freedom in gaining information about the phase $\varphi$ with high accuracy, even when one of the resources is constrained. 

\section{Conclusions} \label{Conclusion}

This paper proposes a quantum metrology scheme to measure an unknown phase $\varphi$ between two pathways of light with an accuracy above the standard quantum limit. Our scheme is based on a laser-driven optical cavity inside an instantaneous quantum feedback loop, as illustrated in Fig.~\ref{Cavity-Prep}. The measurement process includes two main steps. Firstly, during the preparation stage, a continuous laser experiences the phase shift, $\varphi$, before entering the cavity field. Its purpose is to prepare the cavity in a coherent stationary state which depends strongly on this phase. Secondly, during the measurement phase, the cavity experiences only the quantum feedback loop. Whenever the spontaneous emission of a photon is detected, a laser pulse, which does not experience $\varphi$ and provides the reference frame for the proposed phase measurement, is activated and displaces the resonator field in a controlled way. 

In this paper we have assumed that the detector that monitors the cavity during the measurement stage determines its second-order photon correlation function $g^{(2)}(T,0)$. This means, it essentially measures the joint probability for the detection of a photon at the very beginning (at $t=0$) and at the end (at $t=T$) of the measurement stage. As shown in Section \ref{Results2}, this second-order correlation function can be used to determine $\varphi$ with an accuracy $\Delta \varphi$ that scales better than what can be achieved classically according to the standard quantum limit in Eq.~(\ref{Limitsclass}). For the parameters that we consider in this paper, we find that $\Delta \varphi$ scales as $T^{-0.71}$ (c.f.~Eq.~(\ref{Limitsclass4})). If we consider instead the mean number of photons seen by the unknown phase $\varphi$ during the preparation stage as the main resource of our quantum metrology scheme, we find that $\Delta \varphi$ scales as $\left(|\alpha_{\rm ss}|^2 \right)^{-0.65}$ (c.f.~Eq.~(\ref{Limitsclass5})). 

To achieve this quantum enhancement, our metrology scheme uses the temporal correlations of an individual quantum system instead of using multi-partite entanglement. It is worth noticing that subsequent measurements on a single quantum system are in general equivalent to single-shot measurements on multi-partite entangled states. Temporal quantum correlations, which cannot be predicted by a linear master equation, constitute an interesting approach for technological applications \cite{Kok2,Kok3,Ref1,Ref2,Ref3}. As shown in Section \ref{Cavity}, the dynamics of the individual quantum trajectories of the cavity field inside an instantaneous feedback loop is indeed non-linear and depends very strongly on the initial state of the resonator, which encodes the unknown phase $\varphi$ \cite{Masalov,students}. As illustrated in Fig.~\ref{Phase-space}, there is constant stretching and growths of the initially occupied phase space volume. The distance between two different states $|\alpha_1 \rangle$ and $|\alpha_2 \rangle$ which correspond to different $\varphi_1$ and $\varphi_2$ increases rapidly in time. 

The main advantage of the quantum metrology scheme which we propose here is that its experimental realisation is relatively straightforward. As mentioned already above, we do not require highly-entangled many-photon states. Although the proposed scheme requires a relatively good optical cavity, it does not require highly efficient single photon detectors. High-quality optical cavities and relatively fast photon detectors are already available in many laboratories worldwide (see for example Refs.~\cite{Kuhn,Kimble,Laser}). We therefore believe that our quantum metrology scheme will be of significant practical interest until highly-entangled many-photon states become more readily available. \\[0.5cm]
{\em Acknowledgements.} We thank C.~C.~Gerry, P.~A.~Knott, B.~Maybee, F.~Torzewska and G.~Vitiello for stimulating and helpful discussions. Moreover, we acknowledge financial support from the UK EPSRC-funded Oxford Quantum Technology Hub for Networked Quantum Information Technologies NQIT.

\end{document}